**Title**
A Modern Non-SQL Approach to Radiology-Centric Search Engine Design with Clinical Validation

**Authors**
Ningcheng Li, MD, MS; Guy Maresh; Maxwell Cretcher, MD; Khashayar Farsad, MD, PhD; Ramsey Al-Hakim, MD; John Kaufman, MD, FSIR; Judy Gichoya, MD.

**Abstract**
Healthcare data is increasing in size at an unprecedented speed with much attention on big data analysis and Artificial Intelligence application for quality assurance, clinical training, severity triaging, and decision support. Radiology is well-suited for innovation given its intrinsically paired linguistic and visual data. Previous attempts to unlock this information goldmine were encumbered by heterogeneity of human language, proprietary search algorithms, and lack of medicine-specific search performance matrices. We present a de novo process of developing a document-based, secure, efficient, and accurate search engine in the context of Radiology. We assess our implementation of the search engine with comparison to pre-existing manually collected clinical databases used previously for clinical research projects in addition to computational performance benchmarks and survey feedback. By leveraging efficient database architecture, search capability, and clinical thinking, radiologists are at the forefront of harnessing the power of healthcare data.

**Keywords**
Imaging informatics, search engine, NoSQL, document-based database, clinical validation, data-mining

**Introduction**
In an era in which the application of Artificial Intelligence in health care is not only anticipated but being realized (1,2), and deep learning techniques are being applied to datasets with ever increasing size and complexity (3), efficient and accurate access to data is of the uttermost importance. This is relevant in radiology as access to data holds the promise of refining educational content, improving patient care, advancing clinical research, streamlining quality assurance and performance monitoring (4,5). Efficient and accurate access to imaging data further provides vast opportunities for deep learning on the uniquely paired linguistic and visual clinical information, and potentially allowing us to absorb unsuspected features from unsupervised learning algorithms like we did in other disciplines (6). Unfortunately, this access is usually encumbered by heterogeneity of human language, proprietary search algorithms, and lack of medicine-specific search performance matrices.

We aimed to address some of these issues at an institutional level, with specific attention on database architecture, indexing, query processing, user interface, security and clinical validation of a radiology-centric search engine. The result is an adaptable and efficient architecture, allowing improvement of data structure with user feedback and design modification. It is integrated with built-in natural language processing (4,7). Its medical accuracy was validated by comparing the search results to manually collected clinical databases with previously published clinical research results. With widespread application, we hope to establish this framework as the foundation for future deep learning projects in education and clinical research.

**Methods**
*Architecture and Indexing*
Apache Solr (Apache Lucene and Apache Solr, version 7.6.0, https://lucene.apache.org/), a standalone and full-featured search engine class library written in Java was chosen as the backend given document-based database structure, scalability with large datasets, and open-source nature (8,9). Preliminary indexing fields, including "PatientID", "PatientName", "StudyDescription", etc. were defined at the initial construction. Individual radiology reports, treated as "documents", were retrieved from the Oracle databases in batches and converted to JSON type files for consistency prior to upload through the Apache Solr interface. The process of retrieval and upload happens automatically every 20 minutes, every midnight, and every 6 months in order to maintain currency and completeness. Given the unique nature of document-based database with its built-in flexibility, additional indexing fields, such as "PatientDOB" and "ReportUploadDatetime", were included later in the process through user feedback and design modification. Our architecture is summarized in **Figure 1**.

*Query techniques*
A user-input query string was first parsed with a word tokenizer and subsequently stemmed (Natural Language Toolkit, NLTK version 3.4.0, https://www.nltk.org/). If a Boolean operator, including uppercased "AND", "OR", and "NOT" or symbols "+", "-", "|", "&", and "!", were detected, a separate logic processing would be invoked. If a Boolean operator were not detected, a regular search would be assumed with typical stopwords (high-frequency words with little lexical context and distinguishing features, such as "the", "to", and "there") removed. For any query with less than or equal to 4 keywords, all keywords would be mandatory; for queries with between 5 and 9 keywords (inclusive), 20% of the keywords (up to 2) can be optional; for queries with more than or equal to 10 keywords, 30% of the keywords can be optional. Wildcards, denoted by "?" and "*" for one wild card character and multiple wild card characters respectively, were allowed in the regular search. Specific field search used a format of Field_Name:Query_String, e.g. PatientID:123456.

The processed query string was then trialed on several different search patterns: bigram, trigram, and passage search with different allowable word re-arrangement and search field emphasis.

Bigram and trigram searches focus predominantly on patient name, report author name, and report title while passage search focuses on imaging report findings and impressions. Filtering options, including modality, time range, and collapsed results by patient identifier, were applied. Search results were then ranked based on a combined score from the aforementioned searches with a search field weighting system as well as a recency booster (recent imaging reports would be ranked higher). As a final step, display pagination was generated based on the total number of search results, number of results per page, and user-specified page number. Specific checkpoints were set at every stage to prevent malignant manipulation of the search engine.

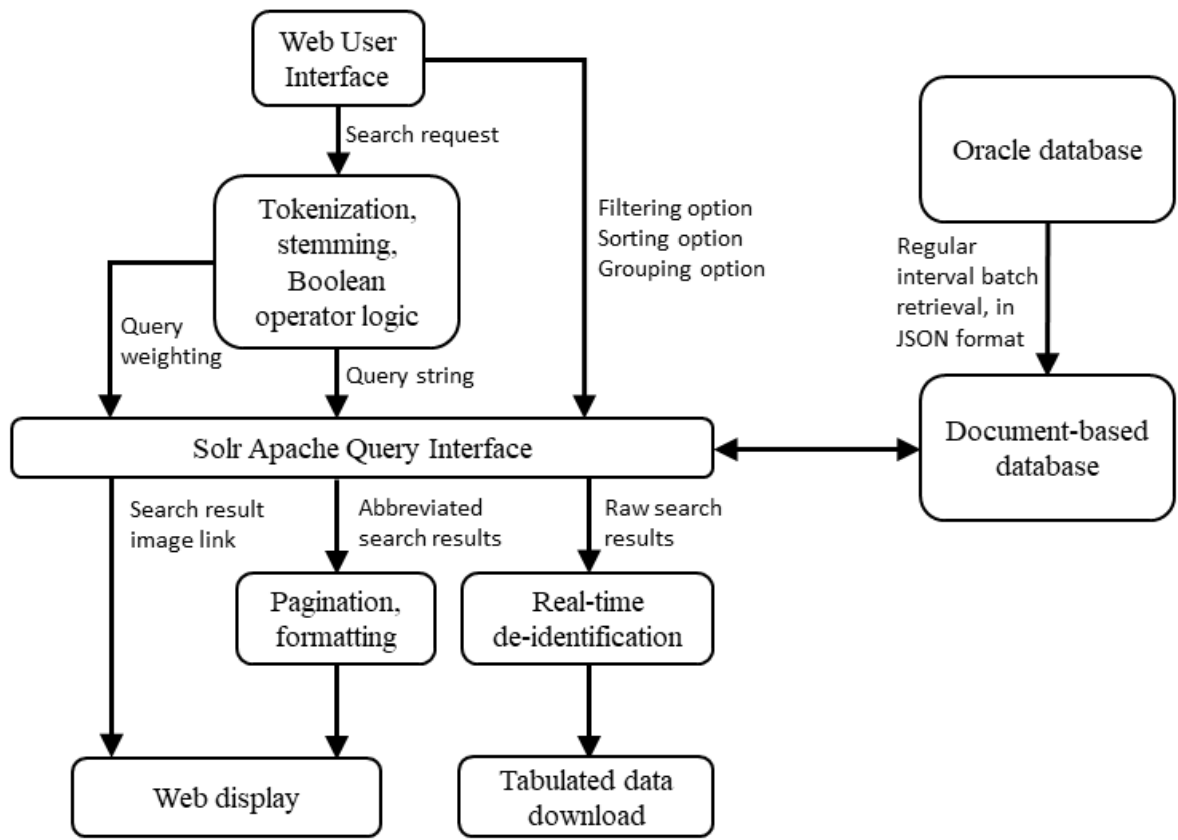

**Figure 1** – Architectural overview of our enterprise radiology search engine.

*HIPAA and Security*
HIPPA authorized disclosure of patients' PHI for specific health care operations including education/training, quality assurance, management, and administration, as well as reviews preparatory to research, without prior IRB approval (10–12). To monitor and prevent inappropriate usage of the search engine, user specific activity on the search engine was logged to create an audit trail to comply with HIPAA regulation based on prior design considerations (13) and would be reviewed if inappropriate usage or HIPAA violation were suspected.

In order to maximize security and minimize the possibility of HIPPA breach from the design aspect, four independent layers of credential verification were staged at each strategic checkpoint. Our enterprise institution secure connection, which can be established on-site via institutional log-in or through VPN, is required for access of the Radiology Search website. A separate credential login is required for actual search capability, with the website staff team managing and monitoring the registration process. An additional credential check, usually requiring the submission of an approved IRB protocol, is required for advanced display of results, including the complete list download of de-identified reports, for ease of conducting research. A final credential login is necessary for the use of Xero Viewer (Agfa-Gevaert Group, Xero, version 8.0.0, https://global.agfahealthcare.com/main/enterprise-imaging/universal-viewer/) for image display.

*Graphical User Interface*
A graphical user interface was designed using a Django (Django Software Foundation, Django, version 2.2, https://www.djangoproject.com/) framework with formatted display of search results and direct linkage to Xero Viewer for radiology images. The search results are retrieved in batches of 10; no more than 10 individual reports are available at the back end at any time point, minimizing the possibility of a large-scale data breach. These batches of 10 search results are compressed into an accordion display, prohibiting direct copying of the search results and further deterring inappropriate or malevolent usage. The accordion can then be expanded individually to present the radiology report and its associated patient information. An icon with a study-specific embedded link to Xero Viewer is displayed alongside each search result for instant viewing of the study images. The graphical user interface was also designed in a mobile friendly fashion with mobile specific display proportions and arrangement of search results (**Figure 2**).

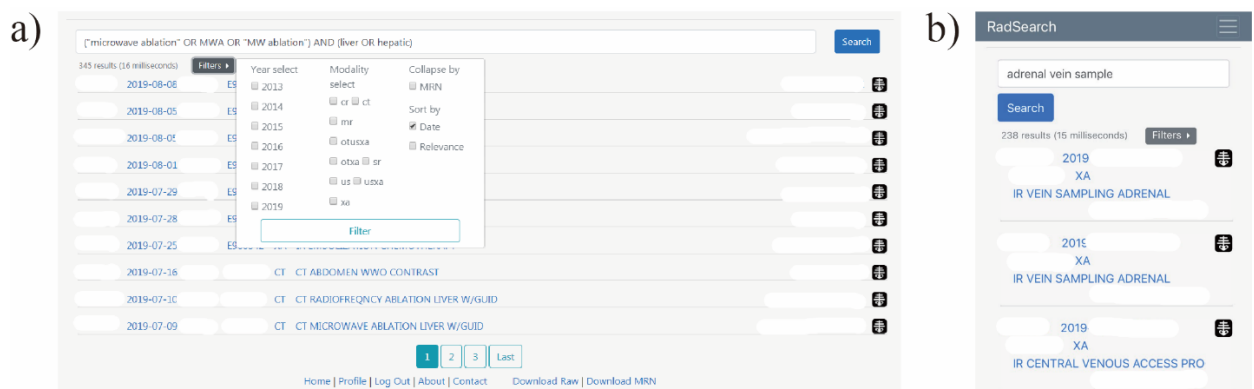

**Figure 2** – Examples of the graphical user interface of the search engine. a) Search results in an accordion display on the desktop version. b) Search results in an accordion display on the mobile version.

*Computational assessment*

Assessment of the search engine's performance in a demanding medical imaging environment is crucial. With an IRB-approved retrospective study protocol, search query data was collected from 49 active users within a 6-month period (December 2018 to May 2019). Search engine response time and number of results for each query were analyzed for basic performance and scalability. Search result page navigation data was also collected. Common search purposes were extrapolated from user feedback and feature requests. Unlike typical web-based commercial search engines, a search engine built specifically for the purpose of medical imaging query and research mandates high standards in sensitivity (inclusion of true imaging description of a certain query) and specificity (exclusion of negative imaging description of a certain query). In order to assess the search engine's clinical accuracy, five pre-existing manually collected clinical databases used previously for clinical research projects were available as the "gold-standard" for comparison (each with a separately approved IRB protocol). Search queries with an increasing number of operators (up to 20) were used for a gradient assessment of the search engine's specificity and sensitivity.

*User survey*

A user survey comprising ten questions was sent out to all registered users assessing their usage of the search engine. The following questions were included: 1) clinical position; 2) search engine use frequency; 3) whether security measures hindered usability; 4) purpose of search; 5) search engine usage location; 6) search operator usage frequency; 7) number of operators used; 8) number of pages navigated; 9) reason(s) for scrolling past the first page of search results; 10) free-form text feedback.

Data was presented as mean ± standard errors of the mean (SEM) and frequency (percentage) for numerical and categorical variables, respectively. Analysis was completed using Mann-Whitney nonparametric t-test and linear regression. A p value less than 0.05 was considered significant.

**Results**

*Search engine performance benchmark*

The Radiology Search Engine went live in December 2018. By May 2019, 3,271,247 imaging reports were indexed and available for search, spanning from year 2002 to 2019. 62 unique users were registered on the Radiology Search Engine and 49 of them had at least one search query. 11,287 searches were performed among the 49 users with an average of $230.3 \pm 83.9$ searches per user. An average number of $34,138 \pm 2,225$ results was returned per query. The search engine response time was clocked at $160.1 \pm 3.9$ milliseconds (**Figure 3**). The response time (in milliseconds) demonstrated a positive relationship with the number of results (linear regression line of $y = 0.0005918x + 139.9$, significantly non-zero slope with a p value < 0.0001).

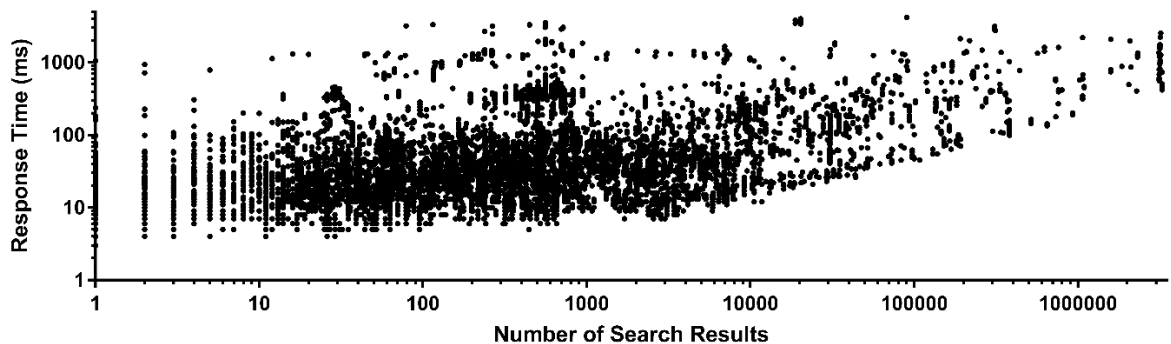

**Figure 3** – Scatter log-log plot showing OHSU Radiology Search Engine's response time versus the number of results obtained from the query.

*Query result navigation pattern*
9,250 out of 11,287 searches (82.0%) resulted in > 0 and ≤ 10,000 query responses. Users navigated to page number 12.6 ± 0.4 on average, with the first 5 pages accounting for 67.6% of all navigation clicks (**Figure 4**). 8,664/9,250 (93.7%) queries had > 10 and ≤ 10,000 results, necessitating multi-page display and potential non-first-page navigation. For these multi-page queries, users navigated through 28.0% ± 0.3% of all available result pages on average (**Figure 4**). 21.7% of the time, users navigated through more than 50% of all available result pages; of all queries with > 20 results [i.e. necessitating pagination of at least 3 pages; 8309/9250 (89.8%) queries qualified], users navigated through more than 50% of all available result pages 19.6% of the time). 3.3% of the time, users navigated through all available, 14.6 ± 3.9 on average, result pages (range: 2-324 pages).

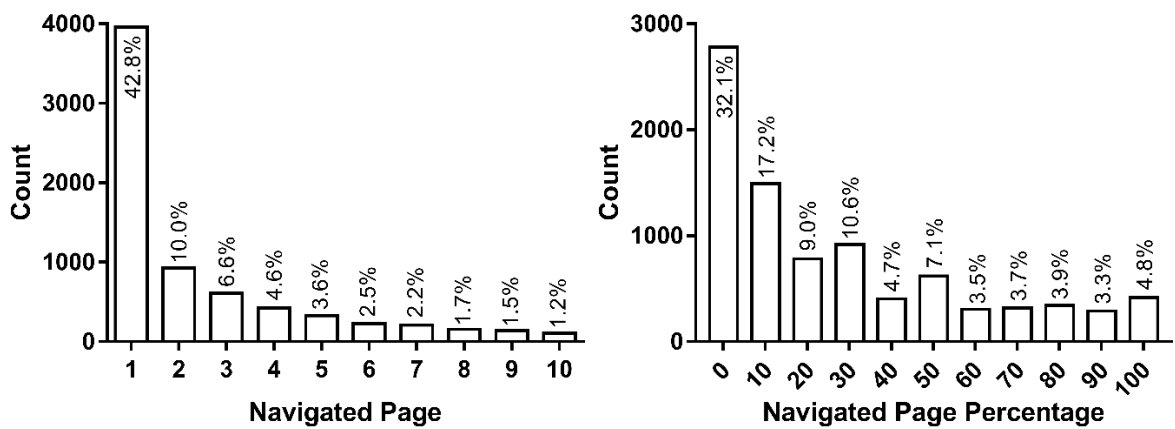

**Figure 4** – Query result navigation pattern. Left: Numeric counts of navigation clicks to a specific result page for all queries with > 0 and ≤ 10,000 results. Right: Percentage of query

result pages that were navigated through by the user for all queries with > 10 and ≤ 10,000 results.

*Search engine sensitivity and specificity*
Five pre-existing, manually collected clinical databases previously used for clinical research were set as the "gold-standard" for assessment of the search engine's sensitivity and specificity. Focuses of these clinical databases included anoxic/hypoxic brain injury, hepatic arterial infusion pump, inferior vena cava filter retrieval, inferior vena cava stent placement, and Palmaz vascular stent. There were a total of 636 patients or on average 127.2 ± 103.1 patients in these five databases. Search queries with an increasing number of operators, up to 20, were utilized to achieve high sensitivity and/or specificity (**Figure 5**). 5/5 (100%) of the search scenarios achieved 100% sensitivity when compared to their respective gold-standard databases. The average specificity at the end of the search was 96.4 ± 2.3% with 2/5 (40%) of the search scenarios achieving 100% specificity. With up to 4 operators, search queries can achieve sensitivity higher than 90% (p = 0.033). With up to 15 operators, search queries can achieve specificity higher than 90% (p = 0.027).

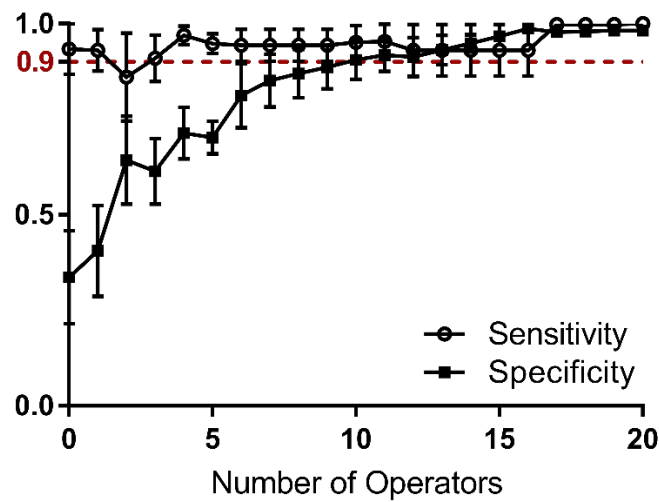

**Figure 5** – Summary data of OHSU Radiology Search Engine results compared with pre-existing manually collected "gold-standard" clinical databases. Sensitivity and specificity were plotted against the number of operators used in the search query string. Error bars denoted standard errors of the mean.

Boolean operator "OR" or equivalent was used in 2/5 (40%) scenarios to achieve better sensitivity (e.g. "anoxic OR hypoxic"). Boolean operator "AND" or equivalent was used in 1/5 (20%) scenarios for a more specific search. Phrasing with double quotations was used in 5/5 (100%) scenarios to denote specific word phrases (e.g. "IVC stent"). Boolean operator "NOT" or equivalent were used in 4/5 (80%) scenarios to achieve better specificity (e.g. "NOT "filter placement"") and accounted for 79.9% ± 10.2% of all operators used in these 4 scenarios.

*User survey*

42 survey responses were collected (67.7%), with 40.5% collected from trainees and 57.1% from faculty members. 33.3% of the users reported usage several times a week. 83.3% reported using the search engine at least once per month. 67.5% of users did not find the extra security measures hindering the search engine's usability. The most common reported purpose of a search was "lecture preparation" at 65.9% followed by "case review preparatory to research", "cases assembly", and "sample report template" at 53.7%, 53.7%, and 43.9%, respectively. 53.7% of the users used the search engine remotely at home while 4.9% used the search engine on a mobile device. 90.0% of users reported using search operators with 45.0% using 1 operators and 40.0% using 2-3 operators. 97.5% of the users reported scrolling past the first page of search results with 25.0% reported scrolling past the $5^{th}$ page, and 15.0% reported scrolling past the $10^{th}$ page. The most common reasons for scrolling past the first page were "exploratory search for multiple examples", "exploratory search for single best representative case", "not seeing the single desired result", "case assembly or procedure tracking requiring multiple cases", and "research requiring multiple cases", at 62.2%, 59.5%, 51.4%, 46.0%, and 32.4%.

**Discussion**

We have described a search architecture that is non-proprietary, non-SQL, secure, efficient, and accurate in the context of medical radiology. Prior efforts at various different institutions have demonstrated a rapid adoption for both clinical and scientific purposes once a clinical search tool became available (13–16). We experienced a similar surge of activity after our institution Radiology Search Engine went live. Usage was strong among both trainees (residents and fellows) and faculty. Junior level trainees tended to use the search engine for "conforming" purposes, including typical description of normal/abnormal imaging features ("anoxic brain injury"), typical diagnoses for specific imaging descriptions ("heterogeneous bone marrow signal"), template reports for a specific body region or indication ("MRI knee"), specific attending style ("Dr. X MRI lumbar spine"), and protocolling for a certain indication ("asymmetric hearing loss MRI"). Senior level trainees focused more on modality/procedure tracking, near-misses ("addendum OR differ"), educational lecture preparation, test preparation, and research. Faculty users focused on lecture preparation, case assembly, reviews preparatory to research, and research with IRB approval. This pattern was similar to previously published results (13). Formal survey feedback and informal discussions with users were overwhelmingly positive. Putative advantages of this system included its fast response time, instant imaging feedback, multi-window viewing allowing efficient screening of multiple cases, and search depth enabled by the extensive Boolean operations as well as stemming.

Our search engine differed in three major aspects compared to prior attempts. It is built upon a document-oriented database structure. Document-oriented database is a specific type of NoSQL or non-structured-query-language, which has an increasing presence in big data and real-time web applications (17). Well-known examples including MongoDB and Couchbase (18–20). In a document-oriented database, all information of a given object (e.g. report findings, report impressions, and report author name of a certain radiology report) is stored within a single instance ("document") in the database, while a separate document might contain different sets of indexable fields (e.g. report dictation time or report update time). This structure differs significantly from the traditional relational databases with predefined categorization schemas and table-like row-based data entries. Document-oriented database offers design simplicity, ability to modify indexing fields at any time as needed, and flexibility in data storage. All of these features are greatly applicable in the design of a medical search engine. As a result, we were able to conceive and develop the Radiology Search Engine, add on indexing fields as we receive feedback from our initial beta test users, and combine more than one radiology report databases.

The Radiology Search Engine is robust, allowing layering parentheses and all the common Boolean operations for complex logic without compromising speed of the search. With an average response time of 160 milliseconds and roughly 6 milliseconds increase with every 10,000 increase in query results, the search engine allows smooth integration into daily clinical and research workflow. It also permits extra time on the backend for further pre- and post-handling if necessary, including natural language processing and advanced filtering/ranking system. Furthermore, it provides a significant edge to deep learning implementation, permitting more training of a classifier within the same timeframe.

Thirdly, this search engine was verified against "gold-standard" databases. To the best of our knowledge, no prior publication on medical search engine design has tested their query results against known, manually collected datasets. Typical commercial search engines placed heavy emphasis on initial "clicks" (the top 5 Google search results accounted for 73.9% (34.4%, 17.3%, 10.4%, 6.9%, and 4.9% respectively) of all clicks (21) and 75% of people never scrolled past the first page (22)). In contrary, accuracy (high specificity) and completeness (high sensitivity) are crucial in a medical context. Our Radiology Search Engine users navigated past the first result page more than half of the time. Users also clicked through a large percentage of available query result pages. Furthermore, some users navigated through all available result pages, including queries with more than 3,000 responses. These behaviors are uncommon in other search environments. And the survey reported reasons for scrolling past the first page demonstrated the importance of completeness in a clinically relevant search engine. With five clinical databases as the "gold-standard", our search engine was able to achieve 100% sensitivity. This is important for research and case assembly as the true positive cases should be included without inherent bias or unexplained exclusion. Indeed, we were able to identify several missed cases and update two of the existing clinical databases in the validation process. The search engine was also able to achieve an average specificity of 96%. This high specificity is beneficial for workflow improvement as it reduces additional time and person-power investment for case screening. Although a higher search specificity is desirable, we found this non-100% specificity a result of human language heterogeneity. In our efforts to achieve higher specificity, we noticed an almost non-deteriorating trend of search sensitivity. This is likely a result of the strategy used (high usage of the Boolean operator "NOT" or equivalent) and the nature of radiology report dictations. For example, a significant number of dictation reports for an initial inferior vena cava filter placement returned as match for inferior vena cava filter removal query. This is due to a recent trend of including recommendations for eventual filter removal during the initial filter placement procedure. Specific wording for such recommendations varied minimally across our institution and these reports were excluded from the search results with three "NOT" operators without decreasing the search sensitivity. Another common scenario involves the communication of a negative finding (e.g. "no evidence of anoxic injury"). No detrimental effect on search sensitivity was observed when we applied a "NOT" operator to remove reports of this nature.

With the appropriate expertise and time, it is feasible to follow the approach outlined in this paper to develop a secure, efficient, and accurate medical search engine de novo. Our design team comprised of three veteran programmers (N.L., G.M., and J.G.) with experience in database design and management, in various stages of their clinical career. The estimated total person-hours prior to beta testing was 400 (240 for front-end development and mobile compatibility, 100 for backend development, 60 for customization, quality assurance, training, regulatory navigation, etc.). An additional 40 person-hours were invested towards the official launch. Maintenance person-hours after official launch were estimated to be 5 per month. In an environment in which such expertise and/or time investment are not available, outsourcing based on specific design concepts described in this paper remains a viable option.

This study and its design concepts are not without limitations. First, the implementation of a medical search engine requires extensive expertise in multiple disciplinaries and regulatory approval. We were extremely fortunate to have the support of our institution's Radiology community and regulatory bodies. Second, a spellchecking and/or spelling suggestion feature was not included. Even though wildcards and stemming are allowed, inadvertent misspelling could return incomplete or empty results. A higher level of natural language processing is desired for better refinement of search results, especially in the cases of pertinent negatives, as evident by the high usage of Boolean operator "NOT" in our validation process. Third, the optional proportion in a long query string and the search field weighting were determined empirically without vigorous recursive optimization. We plan to update these parameters in the future and potentially allow automatic adjustments based on collective user interaction. Fourth, the clinical databases used for validation were pre-existing, published databases with relatively low levels of required search complexity. Larger, more complex searches may be required to fully validate the search engine's accuracy. Lastly, this current implementation is a text-only search engine. We aim to implement a multimodal search engine, as outlined in detail by Pinho (16), to allow concurrent processing of linguistic and pixel data. Several additional important features are in the works. "Real-time anonymization" at the time of query request is a built-in feature. However, permission from HIPPA and IRB regulatory bodies for de-identified non-PHI data download and platform-restricted re-identification key is pending further review. Integration with electronic medical record systems (such as EPIC) and SAP Web Intelligence Business Objects reporting with a Mirth HL7 structure is also under construction.

**Conclusion**
We present a complete process of creating a document-based, secure, efficient, and accurate search engine specifically for radiology. A clinically relevant accuracy matrix was developed for assessment of the search engine's specificity and sensitivity. We believe this search engine will pave the way for future deep learning projects, potentially utilizing this uniquely paired linguistic and visual data in an effective manner. Moreover, we hope that this paper serves as a template for clinically validating radiology data which is an essential substrate for developing meaningful big data/ AI tools to improve our patient outcomes.

**Acknowledgement**
Our work would not have been possible without the support of the Open Source community. We would like to offer our special thanks to Apache Lucene, Apache Solr, Bootstrap, CentOS Linux, Django, NLTK, and Python.

We also give our most sincere thanks to Dr. Bryan Wolf and the rest of our institution radiology community for support and feedback on the Radiology Search Engine.

**References**
1. Minor LB. Harnessing the Power of Data in Health. Stanford Med Heal Trends Rep. 2017;(June):1–18.
2. Langlotz CP. Will Artificial Intelligence Replace Radiologists? Radiol Artif Intell. 2019 May 15;1(3):e190058.
3. Hinton G. Deep Learning—A Technology With the Potential to Transform Health Care. JAMA. 2018 Sep 18;320(11):1101.
4. Cai T, Giannopoulos AA, Yu S, Kelil T, Ripley B, Kumamaru KK, et al. Natural Language Processing Technologies in Radiology Research and Clinical Applications. RadioGraphics. 2016;36(1):176–91.
5. Tang PC. Key Capabilities of an Electronic Health Record System. Washington, D.C.: National Academies Press; 2003.
6. Silver D, Schrittwieser J, Simonyan K, Antonoglou I, Huang A, Guez A, et al. Mastering the game of Go without human knowledge. Nature. 2017 Oct 18;550(7676):354–9.
7. Bird S, Klein E, Loper E. Natural Language Processing with Python. 2nd ed. O'Reilly Media; 2010.

8. Smiley D, Pugh E, Parisa K, Mitchell M. Apache Solr Enterprise Search Server. 3rd ed. Packt Publishing; 2015.
9. Karambelkar HV. Scaling Big Data with Hadoop and Solr. 2nd ed. Packt Publishing; 2015.
10. Use and Disclosure for Treatment, Payment and Health Care Operations. Code of Federal Regulations. Title 45. 164.506. United States of America; 2013.
11. Definition: health care operations. Code of Federal Regulations. Title 45. 164.501. United States of America; 2013.
12. Uses and disclosures for which an authorization or opportunity to agree or object is not required. Code of Federal Regulations. Title 45. 164.512(i)(1)(ii). United States of America; 2013.
13. Erinjeri JP, Picus D, Prior FW, Rubin DA, Koppel P. Development of a Google-Based Search Engine for Data Mining Radiology Reports. J Digit Imaging. 2009;22(4):348–56.
14. Sharpe RE, Sharpe M, Siegel E, Siddiqui K. Utilization of a radiology-centric search engine. J Digit Imaging. 2010;23(2):211–6.
15. De-Arteaga M, Eggel I, Do B, Rubin D, Kahn CEJ, Muller H. Comparing image search behavior in the ARRS GoldMiner search engine and a clinical PACS/RIS. J Biomed Inform. 2015;56:57–64.
16. Pinho E, Godinho T, Valente F, Costa C. A Multimodal Search Engine for Medical Imaging Studies. J Digit Imaging. 2017;30(1):39–48.
17. Andlinger P. RDBMS dominate the database market, but NoSQL systems are catching up. DB-Engines.com; 2013 Nov.
18. Couchbase. Couchbase Under the Hood - An Architectual Overview. Santa Clara, CA 95054, United States; 2019.
19. Couchbase. Couchbase vs. MongoDB$^{TM}$ for Query. Santa Clara, CA 95054, United States; 2019.
20. Couchbase. Couchbase vs. MongoDB$^{TM}$ for Scale-Out and High Availability. Santa Clara, CA 95054, United States; 2019.
21. Google Organic Click-Through Rates (CTR). Advanced Web Ranking. 2019.
22. Google Search Behavior. Search Engine Optimization. 2019.